\documentclass[twocolumn,english,aps,reprint, superscriptaddress,showpacs,longbibliography, showkeys,pra]{revtex4-2}
\usepackage{amsmath,amssymb,bbm,dsfont, mathrsfs,bm,braket,color,graphicx,comment} 
\usepackage[table]{xcolor}
\usepackage{tabularray}
\usepackage[colorlinks,citecolor=blue,urlcolor=blue,linkcolor = blue]{hyperref}
\usepackage[mathscr]{euscript}
\usepackage{enumitem}
\usepackage{dsfont}
\usepackage{amsthm}
\usepackage{nicematrix}
\usepackage{marginnote}
\usepackage{multirow}

\begin{document}
\title{SWAP-less Implementation of Quantum Algorithms}

\author{Berend Klaver}
\affiliation{Institute for Theoretical Physics, University of Innsbruck, A-6020 Innsbruck, Austria}
\affiliation{Parity Quantum Computing GmbH, A-6020 Innsbruck, Austria}

\author{Stefan~M.A.~Rombouts}
\affiliation{Parity Quantum Computing Germany GmbH, 20095 Hamburg, Germany}

\author{Michael Fellner}
\affiliation{Institute for Theoretical Physics, University of Innsbruck, A-6020 Innsbruck, Austria}
\affiliation{Parity Quantum Computing GmbH, A-6020 Innsbruck, Austria}

\author{Anette~Messinger}
\affiliation{Parity Quantum Computing GmbH, A-6020 Innsbruck, Austria}

\author{Kilian Ender}
\affiliation{Institute for Theoretical Physics, University of Innsbruck, A-6020 Innsbruck, Austria}

\author{Katharina Ludwig}
\affiliation{Parity Quantum Computing Germany GmbH, 20095 Hamburg, Germany}

\author{Wolfgang Lechner}
\affiliation{Institute for Theoretical Physics, University of Innsbruck, A-6020 Innsbruck, Austria}
\affiliation{Parity Quantum Computing GmbH, A-6020 Innsbruck, Austria}
\affiliation{Parity Quantum Computing Germany GmbH, 20095 Hamburg, Germany}

\begin{abstract}
We present a formalism based on tracking the flow of parity quantum information to implement algorithms on devices with limited connectivity without qubit overhead, SWAP operations or shuttling. 
Instead, we leverage the fact that entangling gates not only manipulate quantum states but can also be exploited to transport quantum information.
We demonstrate the effectiveness of this method by applying it to the quantum Fourier transform (QFT) and the Quantum Approximate Optimization Algorithm (QAOA) with $n$ qubits. 
This improves upon all state-of-the-art implementations of the QFT on a linear nearest-neighbor architecture, resulting in a total circuit depth of ${5n-3}$ and requiring ${n^2-1}$ CNOT gates. 
For the QAOA, our method outperforms SWAP networks, which are currently the most efficient implementation of the QAOA on a linear architecture.
We further demonstrate the potential to balance qubit count against circuit depth by implementing the QAOA on twice the number of qubits using bi-linear connectivity, which approximately halves the circuit depth.
\end{abstract}

\date{\today}
\maketitle

\section{Introduction}
At the heart of quantum computation is the possibility to subject a quantum system to a controlled sequence of unitary transformations.
Such sequences can be formulated as quantum circuits that concatenate quantum gates operating on the quantum bits (qubits) of the quantum computer.
To be able to execute a given circuit, quantum devices are required to supply gate operations between all qubits that they need to operate on collectively.
Contrary to classical bits, qubits can neither be read out without loss of information nor be copied~\cite{wootters1982_nocloning}, which requires them to function as both memory and computational unit simultaneously.
This ties a unit of quantum information closely to its physical carrier, and the need for operations between arbitrarily distant qubits can arise, which poses a major challenge in realizations of quantum computers~\cite{Henriet_2020, Harrigan2021, cheng2023, Kim2023, Bluvstein_2023}.

In current physical implementations of such quantum devices, gate operations are typically limited to only a small subset of all possible pairs of qubits, which often does not fit the algorithmic connectivity requirement~\cite{Park2023, Holmes2020, Crooks2018, Harrigan2021, dupont2024quantumoptimizationmaximumcut}.
Intrinsically all-to-all connected hardware is challenging to scale~\cite{kielpinski2002architecture}, because physical interactions vanish at large distances.
As an alternative to such long range interactions, one can use either SWAP operations \cite{ogorman2019, Weidenfeller2022} or shuttling of qubits \cite{lekitsch_blueprint, brown_ionshuttle, moses2023} to ``move'' quantum information around.
While SWAP operations keep the qubits in place and exchange their quantum states, shuttling physically moves the qubits themselves.
Using sufficiently many of these operations, arbitrary connectivity can be emulated on most hardware. 
However, within the constraints of current noisy intermediate-scale quantum (NISQ) hardware~\cite{Preskill2018}, both options are costly and compromise computational fidelity.
Furthermore, finding optimal sequences for both SWAP operations and shuttling is computationally hard~\cite{botea2018complexity,Saeedi2011, Qiskit, Sivarajah2021, Martiel2022}.

In this article, we present a formalism of tracking the flow of parity quantum information (i.e., information about relative qubit alignment) and apply it to implement quantum algorithms in hardware with interactions between nearest neighbors without requiring SWAP operations, shuttling, or a qubit overhead.
It is based on an adaptation of the Lechner-Hauke-Zoller (LHZ) architecture~\cite{Lechner2015} and the parity architecture for universal quantum computing~\cite{Fellner2022_prl, Messinger_2023}.
We demonstrate the effectiveness of this method for the quantum Fourier transform (QFT)~\cite{NielsenChuang2011} and the quantum approximate optimization algorithm (QAOA)~\cite{Farhi2014} for all-to-all connected binary Ising models.
Both algorithms are implemented on linear architectures, where we improve gate counts and circuit depth compared to the best-of-class algorithms found in literature.  
By means of parity compilation~\cite{Ender2023}, the approach can be readily extended to QAOA circuits for higher-order spin models or adapted to specific algorithmic needs, for example to a smaller subset of interactions.

The remainder of this paper is organized as follows.
In Sec.~\ref{sec:labels} we introduce parity label tracking, which is the concept underlying our quantum circuit design.
Building on that, we discuss how this tracking of parity information can be exploited to design efficient quantum circuits in Sec.~\ref{sec:inspired_circuits}. 
Subsequently, Sec.~\ref{sec:applications} presents the application of our findings to the QFT and the QAOA.
Finally, we discuss the use of entangling gates other than the CNOT gate in Sec.~\ref{sec:other_entangling_gates}, before we conclude our work in Sec.~\ref{sec:conclusion}.

\section{Parity label tracking}
\label{sec:labels}
Parity codes~\cite{Ender2023, Fellner2022_prl, Messinger_2023, terHoeven2023} encode information of multiple logical qubits onto single physical qubits in order to manipulate it locally. 
If the multi-qubit information is given along the same axis, e.g., $Z_0Z_1Z_2$ or $X_0X_1$ for Pauli operators $X_i$ or $Z_i$, it is referred to as the \textit{parity} of the qubits involved. 
Typically, parity codes focus on localizing parities of a single basis, since these can always be localized at the same time, which in turn also leads to the appearing stabilizers being defined in the same basis. 
In the following we choose to focus on $Z$ parities, where the physical qubits holding multi-qubit $Z$ information are called \textit{parity qubits}, while physical qubits trivially encoding single-qubit $Z$ information are referred to as \textit{base qubits}~\footnote{Note that the base qubits were termed data qubits in previous literature about the parity architecture, e.g., Ref.~\cite{Fellner2022_prl}.}.

\begin{figure}
    \centering
    \includegraphics[width=\columnwidth]{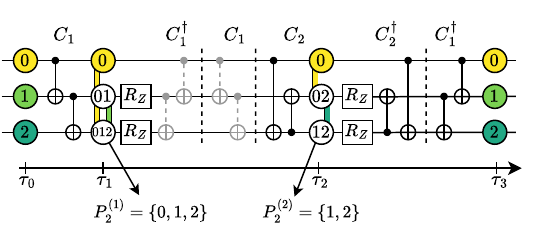}
    \caption{Labeling of qubits demonstrating the effects of CNOT gates on the qubit encoding.
    The labels $P^{(1)}_{2}$ and $P^{(2)}_{2}$ of physical qubit $2$ respectively at time $\tau_{1}$ and $\tau_{2}$ are explicitly shown and used as example in the main text.
    The initial labels $P_j^{(0)}$ directly correspond to individual logical qubit information $\bar{Z}_{j}$ labeled $0,1$ and $2$. 
    Furthermore, the labels change at the targets of the CNOTs according to Eq.~\eqref{eq:symmetric_difference}.
    The colored lines between the qubit vertices represent logical lines, indicating how the corresponding logical $X$ operator is spread out onto physical operators.
    }
    \label{fig:cnot_example}
\end{figure}

In this section, we present the formalism of tracking parity information and show why this is a helpful tool to design efficient quantum algorithms. 
The existing methods to design quantum algorithms that utilize parity information \cite{Wille_2019, Kissinger2019, Nash2020} are typically used in a heuristic manner for a given pair of algorithm and connectivity.
In contrast, we provide quantum circuits for QFT and QAOA on linear architectures for any system size, 
leading to analytical resource estimates.
To this end, we assign a label $P^{(\lambda)}_{a}$ to every physical qubit $a$ at time $\tau_{\lambda}$ and define each qubit label to be the set ${P^{(0)}_{a}=\{a\}}$ at the starting point $\tau_0$ of the label tracking.
During a time step $\lambda$, which is defined as the time period between time $\tau_{\lambda-1}$ and time $\tau_{\lambda}$, the label of a qubit $t$ can be changed by applying the gate ${\rm{CNOT}}_{{c}\rightarrow {t}}$, where the index $c$ denotes the control qubit and $t$ the target qubit of the CNOT gate.
The labels are tracked accordingly as
\begin{align}\label{eq:symmetric_difference}
    P_c^{(\lambda)} &= P^{(\lambda-1)}_{c},\\
    P^{(\lambda)}_{t} &= P^{(\lambda-1)}_{c}\triangle P^{(\lambda-1)}_{t},
\end{align}
where $\triangle$ denotes the symmetric difference between sets, and are illustrated in Fig.~\ref{fig:cnot_example}.
Sets of CNOT gates applied during time step $\lambda$ will be denoted by a unitary operator $C_{\lambda}$ throughout the manuscript.
Hence, a set of CNOT gates between time $\tau_{0}$ and $\tau_{\lambda}$ can be written as a sequence of operators $C_{\lambda}\cdots C_{1}$, giving rise to the time dependent labels.
Using this notation, at every time $\tau_{\lambda}$ we can interpret the initial qubit information to be encoded according to the labels at that time.
In particular, a physical operator $Z_a$ on a qubit with label $P^{(\lambda)}_{a}$ can be understood to act as a logical operation of the form
\begin{equation}
\label{eq:effective_z}
\textstyle\prod_{j\in P^{(\lambda)}_{a}}\bar{Z}_{j}.
\end{equation}
Here and in the following, logical operators are denoted with a bar.
As we will see later, this also translates to general rotations, i.e., exponentials of such operators.
The labels also indicate how to apply a logical $X$ operation on qubit $j$, namely by flipping all the physical qubits that encode $\bar{Z}_{j}$ information and therefore hold index $j$ in their label.
Consequently, the logical operation $\bar{X}_{j}$ can be implemented physically as
\begin{equation}\label{eq:logical_line}
    \textstyle\prod_{b \in \mathcal{L}_{j}^{(\lambda)}} X_{b},
\end{equation}
where $\mathcal{L}_{j}^{(\lambda)}$ contains the indices of all physical qubits which, at time $\tau_\lambda$, contain $j$ in their label.
The sets $\mathcal{L}_{j}^{(\lambda)}$ are also known as \textit{logical lines}~\cite{Fellner2022_prl},  illustrated in Fig.~\ref{fig:cnot_example} by the colored lines connecting the qubit vertices. 
Using Eqs.~\eqref{eq:effective_z} and \eqref{eq:logical_line}, the labels $P_a^{(\lambda)}$ enable us to directly see the effective logical operations implemented by circuits composed of CNOT gates and single-qubit Pauli rotations.

As an example, take the three-qubit circuit which implements all possible multi-qubit $Z$ rotations, shown in Fig.~\ref{fig:cnot_example}.
Therein, let us focus on physical qubit $2$, on which the desired logical operations are $e^{i\alpha\bar{Z}_{0}\bar{Z}_{1}\bar{Z}_{2}}$ and $e^{i\beta \bar{Z}_{1}\bar{Z}_{2}}$.
Applying the circuit $C_{1}=\rm{CNOT}_{1\rightarrow 2}\rm{CNOT}_{0\rightarrow 1}$ changes the label ${P_{2}^{(0)}}=\{2\}$ to ${P_{2}^{(1)}=\{0, 1, 2\}}$, such that the physical single-qubit rotation $e^{i\alpha Z_2}$ effectively applies the logical multi-qubit rotation $e^{i\alpha\bar{Z}_{0}\bar{Z}_{1}\bar{Z}_{2}}$.
Note that applying the inverse sequence $C^{\dagger}_{1}$ (gray in the figure) after the rotation would restore the initial labels. 
This view of encoding, single-qubit rotation and decoding would correspond to a phase gadget~\cite{phase_gadgets_cowtan} of the form
\begin{equation}
\label{eq:conjugation1}
 C^{\dagger}_{1}e^{i\alpha Z_2}C_{1}= e^{i\alpha\bar{Z}_{0}\bar{Z}_{1}\bar{Z}_{2}},
\end{equation}
where the logical rotation on the right-hand side is given by the Clifford-conjugate of the physical rotation on the left-hand side.
As a next step, to implement the second logical operator $e^{i\beta \bar{Z}_{1}\bar{Z}_{2}}$, we need to change a label to $\{1,2\}$. 
Instead of decoding, i.e., restoring the initial labels, we choose to adapt our code by \textit{recoding} from the current set of labels to the required set of labels.
This is achieved by applying another circuit $C_{2}=\rm{CNOT}_{0\rightarrow 2}\rm{CNOT}_{2\rightarrow 1}$, after which the desired rotation can be implemented on qubit $2$.
In this picture, the logical unitaries $\bar{U}$ are applied by alternating sets of Clifford and non-Clifford gates and their effect can be seen via the labels as
\begin{equation}
\label{eq:conjugation2}
C_3 e^{i\beta Z_2} C_{2}e^{i\alpha Z_2}C_{1} = \underbrace{e^{i\beta \prod_{j\in P^{(2)}_{2}}\bar{Z}_{j}}}_{\bar{U}_{2}}\underbrace{e^{i\alpha \prod_{j\in P^{(1)}_{2}}\bar{Z}_{j}}}_{\bar{U}_{1}},
\end{equation}
where $C_{3}=C^{\dagger}_{1}C^{\dagger}_{2}$. 
Note that the final decoding circuit $C_3$ does not necessarily have to be applied when succeeded by only Pauli operator measurements, since these measurement outcomes can be decoded by classical post-processing.
This example demonstrates that tracking the labels is a convenient way to calculate the Clifford-conjugates of physical operators, as one can rewrite Eq.~\eqref{eq:conjugation2} as
\begin{equation}
\label{eq:conjugation3}
(C^{\dagger}_{1}C^{\dagger}_{2} e^{i\beta Z_2} C_{2}C_1)(C^{\dagger}_{1}e^{i\alpha Z_2}C_{1}) = \underbrace{e^{i\beta \bar{Z}_{1}\bar{Z}_{2}}}_{\bar{U}_{2}}\underbrace{e^{i\alpha\bar{Z}_{0}\bar{Z}_{1}\bar{Z}_{2}}}_{\bar{U}_{1}},
\end{equation}
where we make use of the fact that for any operator $M$ and any unitary $C$ the relation
\begin{equation}
    C^{\dagger} e^{i\theta M} C = e^{i \theta C^{\dagger}M C }
\end{equation}
holds.
At time ${\tau_{\lambda}}$, the labeling generalizes the logical operator corresponding to a physical single-qubit $Z$ operator of qubit $a$ to
\begin{equation}
\label{eq:logical_z_operator}
C^{\dagger}_{1}\cdots C_{\lambda}^{\dagger}{Z}_{a}C_{\lambda}\cdots C_{1}=\textstyle\prod_{j\in P^{(\lambda)}_{a}}\bar{Z}_{j}.
\end{equation}
An analog expression generalizes the logical $X$ operators as defined by the logical line $\mathcal{L}_{l}^{(\lambda)}$ (and thus the labels) as 
\begin{equation}
\label{eq:logical_x_operator}
    C^{\dagger}_{1}\cdots C_{\lambda}^{\dagger}\left(\textstyle\prod_{j\in \mathcal{L}_{l}^{(\lambda)}}X_{j}\right)C_{\lambda}\cdots C_{1}=\bar{X}_{l}.
\end{equation}
Additionally, a second set of labels could be tracked or computed from the $Z$ labels for the $X$ basis specifically, which would effectively track the labels of the inverse circuit, leading to a similar expression as Eq.~\eqref{eq:logical_z_operator} for the $X$ basis.
See Ref.~\cite{klaver2025parityflowformalismtracking} for a generalized formalism of parity labels for any Clifford circuit.
The rules for tracking labels for the $X$ basis, $Q^{(\lambda)}_{a}$, follow from the fact that a Hadamard-transform of the CNOT gate effectively interchanges the control and target of this CNOT gate.
Furthermore, the labels can be used to express arbitrary unitary operators $\bar{U}$, by decomposing them into
\begin{equation}
\bar{U}=\prod_\lambda\prod_j \bar{U}^{X}_{\lambda,j}(\theta_j)\bar{U}^{Z}_{\lambda,j}(\phi_j),
\end{equation}
with the unitaries ${\bar{U}^{Z}_{\lambda,j}(\phi)=\exp(i\phi \,\Pi_{k\in P^{(\lambda)}_{j}}\bar{Z}_{k})}$ and ${\bar{U}^{X}_{\lambda,j}(\theta)=\exp(i\theta \,\Pi_{k\in Q^{(\lambda)}_{j}}\bar{X}_{k})}$. 
This decomposition is always possible, since these operators are physically implemented via $X$ and $Z$ rotations and CNOT gates [see Eqs.~\eqref{eq:logical_z_operator} and \eqref{eq:logical_x_operator}], which together form a universal gate set.

\label{sec:spatial_extension}
\begin{figure*}[tb]
\centering
\includegraphics[width=1.0\linewidth]{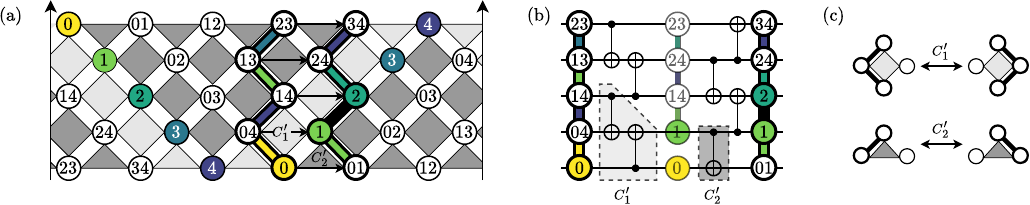}
\caption{(a) A unit cell of the extended LHZ code for $5$ logical qubits in two spatial dimensions.
The large arrows indicate the periodic boundary conditions, i.e., joining the edges indicated by the arrows forms a closed strip with periodic labeling. 
For example, the qubits with labels $\{0, 3\}$, $\{0, 4\}$, $\{1, 3\}$ and $\{1, 4\}$ at the boundaries form a $Z$ stabilizer.
Two different spanning lines are highlighted by the qubits and lines with thick borders, reaching from the top boundary to the bottom boundary. 
In the spatial picture, the colored parts correspond to overlap of a logical line with a spanning line, while in the temporal encoding they correspond to the full logical lines.
The horizontal arrows indicate how the spanning line effectively evolves from left to right by applying the quantum circuit in (b), which changes the qubit labels. 
(b) The quantum circuit that changes the parity labels in the left spanning line to the labels in the right spanning line in (a). 
(c) Pictorial representation of the action of the gates $C'_{1}$ and $C'_{2}$ on the spanning lines.
}
\label{fig:temporal_parity}
\end{figure*}

Although the example in Fig.~\ref{fig:cnot_example} does not have any stabilizers, i.e., it encodes $n$ qubits onto $n$ qubits, it can still be considered a (parity) code.
In order to allow for parallelization or error correction capabilities, redundancy of the encoded information is required, which can be achieved with the help of auxiliary qubits.
If these are initialized in the $+1$ $Z$ eigenstate $|0\rangle$, applying a physical $Z$ gate to these qubits has a trivial effect, as $Z|0\rangle=I|0\rangle$. 
In our formalism, this is represented by giving such qubits an empty set as a label and considering the empty product as an identity, such that the corresponding logical operator is also the identity, i.e., ${\Pi_{j\in\{\}} \bar{Z}_{j}=I}$. 
Accordingly, we call these qubits \textit{empty qubits}, as opposed to \textit{active qubits}, which have a non-empty label. 
An empty qubit can become an active qubit by targeting it with CNOT gates which are controlled by active qubits, such that a non-empty parity label is generated on the previously empty qubit.
Any parity code encoding $n$ logical qubits can be built up from $n$ initial active qubits and any number of additional empty qubits, with a corresponding encoding circuit \cite{NielsenChuang2011}. 
Each added empty qubit adds one stabilizer operator to the code definition. 
At initialization of the empty qubit, the stabilizer is the $Z$ operator of that qubit. 
With the encoding circuit, such single-qubit stabilizers are mapped onto multi-qubit operators, which are represented as gray polygons, as for example shown in Fig.~\ref{fig:temporal_parity}(a).
Note that the tracking method only remains valid as long as no physical operations that violate any stabilizers are performed.

\section{Quantum circuits inspired by the LHZ code}
\label{sec:inspired_circuits}
So far, we have introduced parity labels and their tracking. 
In the following sections, we illustrate the use of the tracking in quantum circuit design based on the LHZ code~\cite{Lechner2015, Fellner2022_prl}, which contains the labels that correspond to the interactions required for the execution of promising quantum algorithms.
The LHZ code emerges when encoding $n$ logical qubits onto additional empty qubits, such that all possible two-body parities appear, by means of tracking the labels throughout a specific encoding circuit~\cite{Fellner2022_prl}. 
This results in $n$ base qubits and $n(n-1)/2$ parity qubits holding all possible single- and two-body parities in their labels.
Changing an empty qubit into a parity qubit induces a new stabilizer on multiple physical qubits, as described in Sec.~\ref{sec:labels}, which are geometrically local on a 2D grid.
The resulting LHZ code for ${n=5}$ qubits appears twice in Fig.~\ref{fig:temporal_parity}(a), shown as qubits connected by dark-gray polygons. 
Both sets of dark-gray polygons correspond to $n(n-1)/2$ stabilizers, which constrain the logical state space. 

\subsection{Spatial extension of the LHZ code}
The LHZ code can be extended in space by positioning two LHZ codes next to each other, where one is turned upside down, such that the light gray stabilizers [see Fig.~\ref{fig:temporal_parity}(a)] fuse the two LHZ codes.
This process can be repeated to create a periodic code, with a period of ${n+1}$ qubits in the horizontal direction, as indicated by the vertical arrows at the left and right boundary of Fig.~\ref{fig:temporal_parity}(a).
An essential property of the extended layout is that any line of physical qubits connecting the top boundary to the bottom boundary contains all logical information, i.e., the information required to construct the complete logical state vector.
In the following, we will denote such sets of qubits as \textit{spanning lines}.
Due to the periodicity of the labels, each spanning line reoccurs after a period of ${n+1}$ qubits.
However, after half a period, i.e., after at most $\lceil\frac{n+1}{2}\rceil$ qubits, every spanning line also appears mirrored around the horizontal axis located exactly in the middle between the top and bottom boundary.
Therefore, the region within half a period in the horizontal direction contains all single-qubit and two-qubit $Z$ parities in its labels. 
This essential feature can be utilized to design a quantum circuit that applies all of these parities, but without the need for any stabilizers. 

\subsection{Temporal parity quantum computing}
\label{sec:temporal}
Given the code defined by the parity labels in Fig.~\ref{fig:temporal_parity}(a) in two spatial dimensions, one can also consider the horizontal axis as a temporal dimension in order to find structured and space-efficient quantum circuits. 
The application of this concept to the QFT and the QAOA is described in Sec.~\ref{sec:applications}.
This translation from a spatial to a temporal encoding reduces the physical qubit requirements to a linear chain of length $n$ to encode $n$ logical qubits, without the need for auxiliary qubits.
However, switching to a temporal encoding implies that only one spanning line is encoded at a time, which means that it is only possible to perform logical operations that solely require the labels currently encoded. 
In the following, we show how to apply the concept of label tracking to change the encoded spanning line, and with that allow for the application of arbitrary logical operations within the parity code.

At the start of the label tracking the qubits in the chain form a spanning line composed solely of base qubits.
Applying a sequence of CNOT gates allows us to change the labels in order to obtain any of the other $n$-qubit spanning lines present in the extended LHZ code.
Let us consider the spanning lines highlighted by the black borders in Fig.~\ref{fig:temporal_parity}(a).
Starting from the labels in the spanning line on the left, the CNOT circuit in Fig.~\ref{fig:temporal_parity}(b) changes the labels to the spanning line on the right.
As an example, the circuit elements $C'_{1}$ and $C'_{2}$ change the label $\{0,4\}$ into $\{1\}$ and label $\{0\}$ into $\{0, 1\}$, respectively. 
When considering this label change in the spatial picture of Fig.~\ref{fig:temporal_parity}(a), the spanning line effectively moves across the stabilizers from left to the right, as indicated by the horizontal arrows.
This picture enables the extended LHZ code to be used as a guide to find the CNOT circuits to change from one spanning line to the other, by expressing the movement across the stabilizers as a CNOT circuit.
If a spanning line contains all of the labels in a given stabilizer except for the label assigned to qubit $a$, a CNOT circuit can be used to deform the spanning line along this stabilizer, i.e., changing the label of one of the overlapping labels into the label of qubit $a$.
This process is illustrated for the spatial picture in Fig.~\ref{fig:temporal_parity}(c) for weight-4 and weight-3 stabilizers.
Since the extended LHZ code exclusively contains such stabilizers, this operation always results in applying one CNOT gate on the boundary or two CNOTs in the bulk, respectively.
The CNOT circuit to change all the labels in a spanning line is composed of two rounds of ${n-1}$ commuting CNOT gates with at most $2$ CNOT targets or controls on the same qubit, as can be seen in Fig.~\ref{fig:temporal_parity}(b).
We assume that a single qubit can only be part of one gate at a time.
For this reason, such a round of CNOT gates is considered to have a circuit depth of $2$.

Using the extended LHZ code as a guide for label tracking is particularly beneficial for the implementation of algorithms that require the interactions specific to the Hamiltonian of a transverse-field Ising model, i.e., two-body $Z$ interactions and single-body $X$ interactions, for which it allows for improved circuit depth and gate count compared to other currently known methods.
The upcoming section introduces explicitly how this is implemented for the QFT and the QAOA.

\section{Applications}
\label{sec:applications}

\subsection{Quantum Fourier Transform}
The QFT is a subroutine for many corner-stone quantum algorithms such as Shor's factoring algorithm. It is given by the unitary operation
\begin{equation}\label{eq:qft}
    U_\text{QFT} = \prod_{i=1}^n \left[ H_i\prod_{j=i+1}^n\text{CP}_{i j}\left(\frac{\pi}{2^{j-i}}\right)\right],
\end{equation}
where $H_i$ denotes a Hadamard gate on qubit $i$. The operation $U_\text{QFT}$ can be represented by a quantum circuit comprising Hadamard and CPhase (CP) gates, where each qubit is entangled with all other qubits [see Eq.~\eqref{eq:qft}].
The CP gate is defined as 
\begin{equation}
    \text{CP}_{jk}(\theta) = \text{diag}\left(1, 1, 1, e^{i\theta}\right),
\end{equation}
where the subscript $jk$ indicates the qubits it acts on.
We are in particular interested in implementing the QFT on a device with linear nearest-neighbor (LNN) connectivity, a problem that was already studied in several works~\cite{Fowler2004lnn, Takahashi2007, Maslov2007lnn, Bhattacharjee2019, Holmes2020, Park2023}, requiring $3n^2/2$ CNOT gates in leading order~\cite{Holmes2020} or introducing a quadratic circuit depth~\cite{Park2023}.
With the above scheme we implement the QFT on an LNN architecture with ${n^2-1}$ CNOT gates and a total circuit depth of ${5n-3}$, which outperforms all other current approaches to the best of our knowledge.

\begin{figure*}
    \centering
    \includegraphics[width=.9\linewidth]{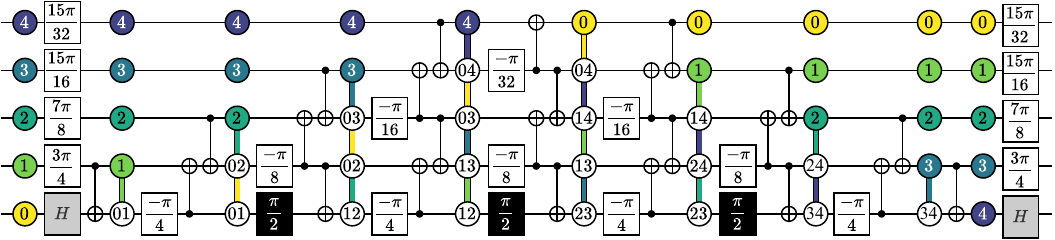}
    \caption{The QFT circuit on a linear chain using the parity label tracking.
    White and black rectangles represent $R_{Z}$ and $R_{X}$ rotations by the indicated angles, respectively, and gray rectangles correspond to Hadamard gates.
    The circles correspond to the physical qubits, and their labels denote the parities of logical qubits they carry.
    At the beginning and the end of the circuit, the physical qubits have a one-to-one correspondence to the logical qubits.
    Note that the algorithm intrinsically reverses the order of qubits.}
    \label{fig:qft_circuit}
\end{figure*}

Our approach starts with a set of qubits holding the initial quantum state and then applies the QFT in the parity architecture as presented in Ref.~\cite{Fellner2022_applications} by iterating through different spanning lines. For that we apply sequences of CNOT gates as depicted in Fig.~\ref{fig:qft_circuit}, which illustrates the corresponding circuit for ${n=5}$ qubits. 
The logical CPhase gates in the QFT circuit are decomposed into three $R_{Z}$ rotations on base- and parity qubits according to 
\begin{equation}\label{eq:cp_decomposition}
    \overline{\text{CP}}_{ij}(\theta) = R_{Z}^{(i)}\left(\frac{\theta}{2}\right)R_{Z}^{(ij)}\left(-\frac{\theta}{2}\right)R_{Z}^{(j)}\left(\frac{\theta}{2}\right),
\end{equation}
and applied on the physical qubits when they appear in the evolving spanning line. In this context, the superscript of the rotation gates denotes the elements of the parity label for a given physical qubit the gate has to be applied on. For the rotations on base qubits, $R_{Z}^{(i)}$ and $R_{Z}^{(j)}$, it is important to consider whether they occur before or after the Hadamard gate on the respective logical qubit, since the base qubit rotations do not commute with the Hadamard gate. 
The Hadamard gates on qubits ${2,\dots, n-1}$ are implemented by decomposing them into logical $R_{Z}$ and $R_{X}$ gates according to
\begin{equation}\label{eq:hadamard_decomposition}
    \bar{H}_{i}=\bar{R}_{Z}^{(i)}\left(\frac{\pi}{2}\right)\bar{R}_{X}^{(i)}\left(\frac{\pi}{2}\right)\bar{R}_{Z}^{(i)}\left(\frac{\pi}{2}\right),
\end{equation}
whereas the gates on qubits $1$ and $n$ can be applied directly, because there are no interfering CPhase gates before and after these gates, respectively. The $R_Z$ rotations on base qubits originating from the CPhase gates before and after the Hadamard gate add up and can be applied as a single gate, together with the $R_Z$ gates from the Hadamard decomposition. For example, the $R_Z(7\pi/8)$ gate on qubit 2 (see Fig.~\ref{fig:qft_circuit}) is composed from
\begin{itemize}
    \item $R_Z(\pi/2)$ belonging to the Hadamard gate $H_2$, 
    \item $R_{Z}(\pi/4)$ from $\text{CP}_{12}(\pi/2)$ [see Eq.~\eqref{eq:cp_decomposition}] and 
    \item $R_Z(\pi/8)$ from $\text{CP}_{02}(\pi/4)$.
\end{itemize}
The $R_X$ part of the Hadamard gate on qubit $i$ is done at a moment in time, where the corresponding logical line only spans one qubit and therefore results in a physical single-qubit rotation (black operations in Fig.~\ref{fig:qft_circuit}).

A resource comparison of different linear QFT implementations is given in Tab.~\ref{tab:qft_resources}. Here and in the following section, we use the gate set $\{R_Z, R_X, H, \text{CNOT}\}$ for comparing gate counts and circuit depth.
Remarkably, our approach surpasses the direct QFT implementation, resulting from Eq.~\eqref{eq:qft} by replacing the CPhase gates by 2 CNOT gates and a $R_Z$ rotation, even when compared to an implementation on an all-to-all connected device in terms of single-qubit-gate counts and total circuit depth.

When using previous approaches for realizing the QFT, it may be necessary to reverse the qubit order afterwards if the QFT is not at the end of the total quantum circuit~\cite{NielsenChuang2011}.
This requires $3\lfloor \frac{n}{2}\rfloor$ CNOT gates and a single time step on an all-to-all connected device and $3n(n-1)/2$ CNOT gates and $3n$ time steps on an architecture with LNN connectivity, where the factor of 3 originates from decomposing SWAP gates into CNOT gates.
As our approach intrinsically reverses the order of qubits throughout the algorithm execution (see Fig.~\ref{fig:qft_circuit}), we eliminate another quadratic amount of CNOT gates on top of the savings in the actual QFT circuit, when taking the correct ordering of qubits into account.

\begin{table}[t!]
\centering
\resizebox{\columnwidth}{!}{
\begin{tblr}{
colspec={|l|cccc|}, rowsep=1.5pt, colsep=2.5pt, hline{1,2,5,8}
}
Resource   & Parity & Ref.~\cite{Park2023} & Ref.~\cite{Holmes2020} & all-to-all~\cite{Holmes2020}\\ 
\#CNOTs    & $n^2-1$ & $n^2+n-4$ & $\frac{3}{2}n^2-\frac{3}{2}n$ & $n^2-n$ \\
\#SQ gates  &$\frac{1}{2}n^2+\frac{5}{2}n-2$ & $\frac{1}{2}n^2+\frac{5}{2}n-2$ & $n^2$& $n^2$\\
total gates & $\frac{3}{2}n^2+\frac{5}{2}n-3$ & $\frac{3}{2}n^2+\frac{7}{2}n-6$ & $\frac{5}{2}n^2-\frac{3}{2}n$ & $2n^2-n$\\
CNOT depth & $4n-4$ & $n^2+n-4$ & $6n-9$ & $4n-6$ \\
SQ depth   & $n+1$ 
& $2n+1$ & $4n-4 $&  $4n-4$ \\ 
total depth & $5n-3$ & $n^2+3n-3$ & $10n-13$& $8n-10$
\end{tblr}
}
\caption{Comparison of the required resources for the QFT on $n$ qubits in a linear chain using different approaches, and additionally to an all-to-all connected system.
The gate set used comprises CNOT gates as well as single-qubit (SQ) $H$, $R_X$ and $R_Z$ gates. Note that, if SQ gates are not executed in parallel with CNOT gates in our approach, there are $2n-1$ steps with solely SQ gates.
In that case, the total depth increases to $6n-5$.}
\label{tab:qft_resources}
\end{table}

\subsection{The Quantum Approximate Optimization Algorithm}
The QAOA was introduced by E.~Farhi~et~al.~\cite{Farhi2014} and is considered a promising candidate to show quantum advantage for optimization problems in the NISQ-era~\cite{Preskill2018}, especially after many variations have been introduced~\cite{Hadfield2019, Blekos2023}.
It follows a hybrid quantum-classical approach to tackle combinatorial optimization problems, encoded in an Ising-like problem Hamiltonian
\begin{equation}
    {H}_P=\sum_{k=1}^n\sum_{j<k} J_{jk}\bar{Z}_j \bar{Z}_k+\sum_{j=1}^n h_j \bar{Z}_j
\end{equation}
on $n$ qubits, where $J_{jk}$ are coupling strengths between qubits and $h_j$ denote local fields.
In its original formulation, the QAOA prepares a candidate state 
\begin{equation}
    \label{eq:qaoa_state}
    \ket{\psi(\bm{\beta}, \bm{\gamma})} = \prod_{j=1}^p U_{X}(\beta_j) U_{P}(\gamma_j)  \ket{+}^{\otimes n},
\end{equation}
where $U_{X/P}(\alpha)=e^{i\alpha {H}_{X/P}}$ and ${{H}_X=\sum_{j=1}^n \bar{X}_j}$. This is used to estimate the cost function
\begin{equation}
E(\bm{\beta},\bm{\gamma})=\braket{\psi(\bm{\beta}, \bm{\gamma})|H_\text{P}|\psi(\bm{\beta}, \bm{\gamma})}.
\end{equation}
In order to approximate the ground state of $H_{P}$, this is then optimized by varying the $2p$ parameters ${\bm{\beta}=(\beta_1, \dots, \beta_p)}$ and ${\bm{\gamma}=(\gamma_1, \dots, \gamma_p)}$ in a quantum-classical feedback loop, seeking configurations which minimize the energy expectation value.

\subsubsection{The QAOA on a linear chain}\label{sec:linear_qaoa}
To perform the QAOA on a chain with linear nearest-neighbor (LNN) connectivity, typically SWAP networks~\cite{Kivlichan2018, Babbush2018, ogorman2019, Hashim2022, Wang2023} are used to ensure connectivity between all interacting qubits throughout the algorithm. 
If all-to-all connectivity is required, the optimal SWAP network results in an overhead of $n^2/2$ CNOT gates in addition to the CNOT gates necessary to implement the interaction terms, resulting in a total number of required CNOT gates of $3n^2/2$ in leading order~\cite{Crooks2018}, while maintaining a linear circuit depth.

\begin{figure*}
\centering
\includegraphics[width=.9\linewidth]{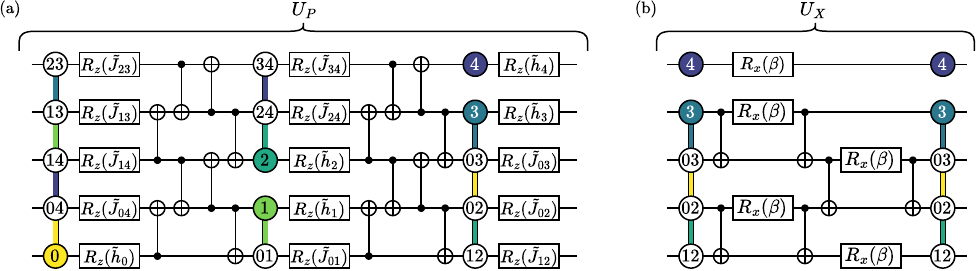}
\caption{(a) The quantum circuit to implement the problem unitary $U_P$ of the QAOA with parameter $\gamma$ by means of the time-encoded parity layout, where the angles are defined as ${\Tilde{J}_{ij}=\gamma {J}_{ij}}$ and ${\Tilde{h}_{i}=\gamma {h}_{i}}$.
The circuit propagates through the different spanning lines as indicated by the changing labels and logical lines (colored lines in the figure).
(b) The quantum circuit to implement the mixer unitary $U_X$ with parameter $\beta$ of the QAOA.
This circuit does not change the spanning line and requires ${2(n-2)}$ CNOT gates during $4$ time steps and $n$ single-qubit gates in $2$ time steps.}
\label{fig:qaoa_circuit}
\end{figure*}

In the temporal parity encoding, $U_P$ is implemented by sequentially transforming an initial zig-zag shaped spanning line to ${\lceil (n-1)/2\rceil}$ other spanning lines and applying single-qubit $Z$ rotations in between each transformation.
The transformation between two zig-zag shaped spanning lines is illustrated in Fig.~\ref{fig:temporal_parity}(a) by the qubits and lines with the thick borders and it is implemented via the circuit design in Fig.~\ref{fig:temporal_parity}(b). 
This method corresponds to covering all the labels present in half a period of the extended LHZ code~\footnote{For an even number of qubits, the last spanning line only needs to differ from the previous one in half of the qubits and its creation only requires ${n-1}$ CNOT gates in 2 time steps. Therefore, the final numbers are independent of the parity of $n$.}. 
By construction, these spanning lines cover all necessary (and, in this case, all possible) labels ${\{j,k\}}$, which allows for the execution of the desired logical operations $\exp(-i \gamma  J_{jk} \bar{Z}_j\bar{Z}_k)$ as physical single-qubit rotations $R_{Z}$ according to Eq.~\eqref{eq:logical_z_operator}.
This implementation of $U_{P}$ is shown in Fig.~\ref{fig:qaoa_circuit}(a) and results in a CNOT depth of ${2n-2}$ and a CNOT gate count of ${(n-1)^{2}}$ (cf.~Sec.~\ref{sec:temporal}).

The unitary $U_X$ can be applied using logical $R_{X}(\beta_{j})$ gates along the logical lines as defined in Eq.~\eqref{eq:logical_x_operator}.
The logical lines at a given time always span at most two neighboring qubits and a spanning line always has at least one localized logical $X$ operator (i.e., logical line of length 1) on a boundary.
Thus, applying $U_X$ requires a constant two-qubit gate depth of ${4}$ and at most ${2n-2}$ CNOT gates.
The circuit diagram for implementing $U_X$ on an exemplary spanning line is given in Fig.~\ref{fig:qaoa_circuit}(b).

In total, our approach exhibits a CNOT gate count of $n^2-1$ and with that outperforms the SWAP network by a factor of $3/2$ in leading order.
Furthermore, it results in a shorter circuit depth while the number of required single-qubit rotations remains unchanged.
The numbers for gate counts and circuit depth in terms of the universal gate set $\{R_Z, R_X, H, \text{CNOT}\}$ are given in Tab.~\ref{tab:qaoa_resources}.
These numbers also hold for the trotterized time evolution of the transverse-field Ising Hamiltonian, as there is a direct relation to the QAOA.

Note that we have not taken into account the initialization of the logical state $\ket{+}^{\otimes n}$ in our discussion. However, this state corresponds to all qubits in the spanning line being in the state $\ket{+}$. Therefore, it can be prepared by simply applying physical Hadamard gates on all physical qubits prepared in the $|0\rangle$ state.

Remarkably, also on two-dimensional grids, no strategies have been found that outperform the linear SWAP strategy~\cite{Weidenfeller2022}.
Consequently, our strategy reduces the resource requirements for QAOA not only on an LNN chain, but also on devices with 2D connectivity.

\begin{table}
\resizebox{\columnwidth}{!}{
\begin{tblr}{
colspec={l|l|ccc|}, rowsep=1.5pt, colsep=2.5pt,
hline{2,6,12},
hline{1} = {2-Z}{solid},
vline{1} = {2-Z}{solid},
cell{2}{1}={r=4}{c},
cell{6}{1}={r=6}{c},
cell{12}{1}={r=6}{c},
}
& Resource        & Linear Parity      & Ladder-Parity & Ref.~\cite{Crooks2018} \\ 
$U_P$ 
& \#CNOTs         & ${n^{2}}-2n+1$        & ${n^{2}}-n+1$         & ${\frac{3}{2}n^{2}}-\frac{5}{2}n+1$ \\
& CNOT depth      & $2n-2$              & $2\lceil\frac{n}{2}\rceil$           & $3n-2$ \\
& \#SQ gates      & $n(n+1)/2 $         & $n(n+1)/2$    & $n(n-1)/2$            \\
& SQ depth        & $\lceil (n+1)/2\rceil$  & $\lceil n/4\rceil$& $n$  \\ 
$U_X$  
& \#CNOTs         & $2n-2$              & $2n-2$        & $0$  \\
& CNOT depth      & $4$                 & $4$           & $0$  \\
& \#SQ gates      & $n$                 & $2n$         & $n$  \\
& SQ depth        & $2$                 & $3$           & $1$  \\
& \# measurements & $0$                 & $n$           & $0$\\
& meas. depth     & 0                   & 1             & 0\\ 
$U_XU_P$ 
& \#CNOTs         & $n^{2}-1$           & $n^2+n-1$& $\frac{3}{2}n^{2}-\frac{5}{2}n+1$ \\
& CNOT depth      & $2n+2$              & $2\lceil\frac{n}{2}\rceil+4$& $3n-2$  \\
& \#SQ gates      & $n(n+3)/2$ & $n(n+5)/2$& $n(n+1)/2$    \\
& SQ depth        & $\lceil (n+1)/2\rceil+2$& $\lceil n/4\rceil + 3$     & $n+1$  \\
& \# measurements & $0$                 & $n$           &$0$\\
& meas. depth& 0& 1&0\\
\hline
\end{tblr}
}
\caption{Measurement, CNOT and single-qubit (SQ) depth and gate counts  for the building blocks of the QAOA for an all-to-all connected problem graph of size $n$ on a device with LNN or ladder connectivity. 
The used gate set comprises CNOT gates as well as single-qubit $H$, $R_X$ and $R_Z$ gates.
Note that in Ref.~\cite{Crooks2018} no local field terms are included. 
Since our presented protocol includes local field terms, the same circuit could also be used to implement the QAOA for a problem of size ${n+1}$ without local field terms.}
\label{tab:qaoa_resources}
\end{table}

\subsubsection{The QAOA on a ladder}
If $2n$ physical qubits are available and arranged in a ladder geometry (i.e., a $2\times n$ grid with nearest-neighbor connectivity), then both rails of the ladder can be used to progress to new spanning lines in parallel, by using the circuit in Fig.~\ref{fig:qaoa_circuit}(a) independently on each rail.
This allows us to reach all required qubit labels 
in half the circuit depth compared to the approach presented in Sec.~\ref{sec:linear_qaoa}.
The gate depth of the full QAOA circuit is then reduced to that of an all-to-all connected system \cite{dreier2025connectivityawaresynthesisquantumalgorithms}, at the cost of doubling the qubit number and gate count, and introducing a linear number of single-qubit mid-circuit measurements.
For the use of measurements in parity code deformations we refer the reader to Refs.~\cite{Messinger_2023, PhysRevLett.132.220602}.

\begin{figure}\includegraphics[width=\columnwidth]{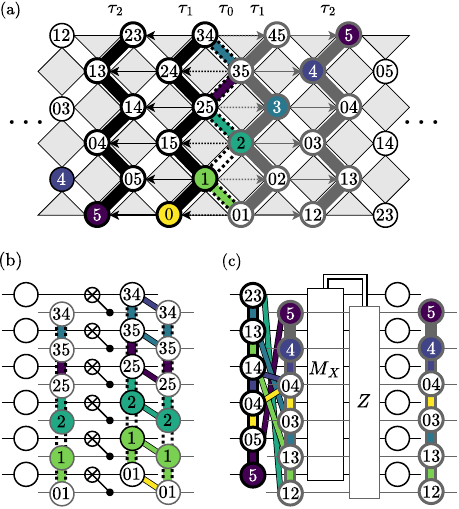}
    \caption{(a) The effective evolution of two spanning lines in opposite directions for applying $U_{P}$ on a ladder architecture for ${n=6}$. The logical lines with dotted borders indicate the initial spanning line ($\tau_0)$. (b) The initialization of a second spanning line from a first spanning line by using empty qubits and $n$ CNOT gates. (c) The two spanning lines after implementing $U_{P}$, where the logical lines are delocalized. In order to allow easier implementation of $U_X$, the encoding is reduced to a single spanning line again, by decoding the second spanning line via $X$-basis measurements and conditional $Z$ gates.
    }
    \label{fig:ladder}
\end{figure}

To allow for parallel evolution of two logical lines, we begin by assuming an arbitrary spanning line on the first rail and empty qubits on the second. 
The second rail is initialized by ``copying'' the labels of the spanning line from one rail to the other rail, using the CNOT circuit shown in Fig.~\ref{fig:ladder}(b). 
This circuit applies $n$ CNOT gates in parallel, where each CNOT gate is controlled by one qubit in the first rail and targets onto its neighboring empty qubit on the second rail of the ladder.
As a result, we obtain two spanning lines with identical labels.

Then, we successively progress the two spanning lines in opposite directions by applying the quantum circuit from Fig.~\ref{fig:qaoa_circuit}(a) to each rail, as depicted in Fig.~\ref{fig:ladder}(a). 
For one of the spanning lines, the targets and controls of each CNOT gate are switched, which reverses the direction of progression for this spanning line. 
Note that for some $n$, it can be beneficial to initially progress each rail by half a step [see transition from $\tau_0$ to $\tau_1$ in Fig.~\ref{fig:ladder}(a)] before applying the first $Z$ rotations in order to reach more parities in that round already.
This approach requires $n$ CNOT gates for the ``copying'' and  $(n-1)^{2}$ CNOT gates for the progression, with a total circuit depth of $2\lceil\frac{n}{2}\rceil$ until all parities have been reached at some time step. 

After applying all interactions in $U_{P}$, the logical lines are delocalized as depicted in Fig.~\ref{fig:ladder}(c). 
To relocalize them such that the logical lines occur along two neighboring qubits on a single rail, we measure all qubits in one rail in the $X$ basis and apply conditional-$Z$ gates to the qubits on the other rail. 
Such a protocol has the same effect as the decoding CNOT sequence required to remove all the labels from the first rail (see \cite{Messinger_2023} for the necessary classical corrections).
This step effectively decodes one rail of the ladder, such that no redundant information remains.
The decoding process adds $n$ qubit measurements and $2n$ single-qubit rotations to the resource requirements of $U_{X}$.
After decoding, $U_{X}$ can be applied on the single remaining spanning line using the same circuit components and therefore the same resource requirements as in Sec.~\ref{sec:linear_qaoa}. 
With that, one QAOA round is finished and we can, if desired, continue by copying the labels of the current spanning line to the second ladder rail again and starting a new round from there.
The overall resource requirements for this method are given in Tab.~\ref{tab:qaoa_resources}.

\section{The use of other entangling gates}
\label{sec:other_entangling_gates}
For devices where the CNOT gate is not a native gate, one can reformulate the above-presented circuits 
by expressing a CNOT gate as a controlled-$Z$ gate or a controlled-phase gate preceded and followed by a Hadamard gate or a $Y$ rotation on the target qubit.
The resource counts from Tables \ref{tab:qft_resources} and \ref{tab:qaoa_resources} still hold up in terms of entangling-gate counts and depths.

If the iSWAP gate is a native gate~\cite{mckay2016iswap}, then we can use the fact that it can be expressed as a sequence of two back-to-back CNOT gates 
(with target and control qubits exchanged from one gate to the other) supplemented with single-qubit gates on both qubits. 
Our circuits display back-to-back CNOT gates in most steps, allowing for a substitution of such pairs by single iSWAP gates.

It should be noted that none of these algorithm implementations require the use of SWAP gates, even for all-to-all connected problem graphs on qubit topologies with minimal connectivity. 
This is made possible by taking advantage of the same operation that implements the interactions from the problem graph, to move quantum information through the connectivity graph.
For a SWAP-based approach to outperform our circuits, the SWAP operation has to be at least as efficient as a single iSWAP gate or two CNOT gates,
assuming that the single-qubit gates have minimal overhead compared to two-qubit gates.
This is unlikely to be the case if the SWAP gate is not a native gate of the quantum computer,
given that a single SWAP gate is equivalent to a sequence of three back-to-back CNOT gates.

\section{Conclusion and Outlook}\label{sec:conclusion}
In this work, we have formalized the tracking of parity labels and demonstrated its utility in quantum circuit design.
For the QFT on a linear chain, we find lower resource requirements in terms of two-qubit gate count and depth for the gate set $\{R_Z, R_X, H, \text{CNOT}\}$ with respect to various other published works and even the same resource requirements as for an all-to-all connected system in leading order.
Our scheme reduces the two-qubit gate count and depth of the QAOA on an LNN architecture by a factor of $2/3$ in leading order compared to previously published methods. 
Moreover, our implementation of the QAOA on a ladder architecture, reducing circuit depth by a factor of two without increasing the gate count in leading order, trades off qubit count and mid-circuit measurements for circuit depth.
These advancements highlight a promising approach for implementing near-term quantum algorithms on NISQ devices, where minimizing gate count and circuit depth is crucial. 
This approach has also been investigated beyond the LNN architecture, where it can exploit the additional connectivity to further reduce two-qubit gate counts~\cite{dreier2025connectivityawaresynthesisquantumalgorithms}.
Further research may involve the development of a compilation tool to leverage these findings for algorithms that do not require all-to-all connectivity.
Additionally, it would be valuable to investigate the actual improvements in algorithmic performance and the reliability of quantum circuit outputs on real quantum devices.
Building on our analytical evaluation of the necessary resources in terms of CNOT gate counts and depth, we anticipate significant improvements not only within the theoretical framework but also in a practical experimental setting, e.g., when subjected to quantum noise.  
Finally, we note that the tracking approach can be generalized to include other Clifford gates and Pauli rotations, related to a concept known as Clifford tableaus~\cite{AaronsonGottesman2004} and flow tableaus~\cite{klaver2025parityflowformalismtracking}. 

\section*{Acknowledgements}
The authors thank P.~Aumann, M.~Lanthaler and G.~B.~Mbeng for valuable comments and discussions.
This study was supported by the Austrian Research Promotion Agency (FFG Project No. FO999909249, FFG Basisprogramm).
This research was funded in part by the Austrian Science Fund (FWF) under Grant-DOI 10.55776/F71 and Grant-DOI 10.55776/Y1067.
This project was funded within the QuantERA II Programme that has received funding from the European Union’s Horizon 2020 research and innovation programme under Grant Agreement No. 101017733. 
For the purpose of open access, the author has applied a CC BY public copyright license to any Author Accepted Manuscript version arising from this submission.
We acknowledge the support of the Federal Ministry of Research, Technology, and Space (BMFTR) within the framework of the program
“Quantum technologies - from basic research to market”, QSOLID, under Grant No. 13N16154.

%

\end{document}